\documentclass[12pt,preprint]{aastex}

\newcommand{\etal}{et al.~}

\shorttitle{M 33 RADIO-SELECTED OBJECTS}
\shortauthors{B{\sc uckalew} et al.}

\begin{document}

\title{UNDERSTANDING RADIO-SELECTED THERMAL SOURCES IN M 33: ULTRAVIOLET, OPTICAL, NEAR-INFRARED, {\it SPITZER} MID-INFRARED, AND RADIO OBSERVATIONS}

\author{B{\sc rent} A.\ B{\sc uckalew}\altaffilmark{1}, H{\sc enry} A.\ K{\sc obulnicky}\altaffilmark{1}, J{\sc onathan} M.\ D{\sc arnel}\altaffilmark{1}}
\altaffiltext{1}{Department of Physics \& Astronomy, University of Wyoming, Laramie, WY  82071, {\tt mrk1236@uwyo.edu}, {\tt chipk@uwyo.edu}, and {\tt jdarnel@uwyo.edu}}
\and
\author{E{\sc lisha} P{\sc olomski}\altaffilmark{2}, R{\sc obert} D.\ G{\sc ehrz}\altaffilmark{2}, R{\sc oberta} M.\ H{\sc umphreys}\altaffilmark{2}, C{\sc harles} E.\ W{\sc oodward}\altaffilmark{2}}
\altaffiltext{2}{Department of Astronomy, University of Minnesota, Minneapolis, MN 55455, {\tt elwood@astro.umn.edu}, {\tt gehrz@astro.umn.edu}, {\tt chelsea@astro.umn.edu}, and {\tt roberta@astro.umn.edu}}
\and 
\author{J{\sc oannah} L.\ H{\sc inz}\altaffilmark{3}, C.\ W.\ E{\sc ngelbracht}\altaffilmark{3}, K{\sc arl} D.\ G{\sc ordon}\altaffilmark{3}, K.\ M{\sc isselt}\altaffilmark{3}, P.\ G.\ P{\sc \'{e}rez}-G{\sc onz\'{a}lez}\altaffilmark{3}, G{\sc eorge} H.\ R{\sc ieke}\altaffilmark{3}}
\altaffiltext{3}{Steward Observatory, University of Arizona, 933 North Cherry Avenue, Tucson, AZ 85721,  {\tt jhinz@as.arizona.edu}, {\tt cengelbracht@as.arizona.edu}, {\tt kgordon@as.arizona.edu}, {\tt kmisselt@as.arizona.edu}, {\tt pgperez@as.arizona.edu}, {\tt grieke@as.arizona.edu}}
\and
\author{S.\ P.\ W{\sc illner}\altaffilmark{4}, M.\ L.\ N.\ A{\sc shby}\altaffilmark{4}, P.\ B{\sc armby}\altaffilmark{4}, M.\ A.\ P{\sc ahre}\altaffilmark{4}}
\altaffiltext{4}{Harvard-Smithsonian Center for Astrophysics, 60 Garden Street, Cambridge, MA 02138, {\tt swillner@cfa.harvard.edu}, {\tt mashby@cfa.harvard.edu}, {\tt pbarmby@cfa.harvard.edu}, {\tt mpahre@cfa.harvard.edu} }
\and 
\author{T.\ L.\ R{\sc oellig}\altaffilmark{6}}
\altaffiltext{6}{NASA Ames Research Center, MS 245-6, Moffett Field, CA 94035-1000 {\tt thomas.l.roellig@nasa.gov}}
\and
\author{N{\sc ick} D{\sc evereux}\altaffilmark{8}}
\altaffiltext{8}{Department of Physics, Embry-Riddle Aeronautical University, Prescott, AZ 86301 {\tt devereux@erau.edu}}
\and
\author{J{\sc acco} T{\sc h}.\ V{\sc an} L{\sc oon}\altaffilmark{5}}
\altaffiltext{5}{Astrophysics Group, School of Physical and Geographical Sciences, Keele University, Staffordshire ST5 5BG, UK {\tt jacco@astro.keele.ac.uk}}
\and 
\author{B.\ B{\sc randl}\altaffilmark{7}}
\altaffiltext{7}{Leiden Observatory, PO Box 9513, 2300 RA Leiden, The Netherlands {\tt brandl@isc.astro.cornell.edu}}

\begin{abstract}
We present ultraviolet, optical, near-infrared, {\it Spitzer} mid-infrared, and radio images of 14 radio-selected objects in M 33.  These objects are thought to represent the youngest phase of star cluster formation.  We have detected the majority of cluster candidates in M 33 at all wavelengths.  From the near-IR images, we derived ages 2--10 Myr, $K_S$-band extinctions ($A_{K_S}$) of 0--1 mag, and stellar masses of 10$^3$--10$^{4}$ $M_{\odot}$.  
We have generated spectral energy distributions (SEDs) of each cluster from 0.1 $\mu$m to 160 $\mu$m.  From these SEDs, we have modeled the dust emission around these star clusters to determine the dust masses (1--10$^3$ $M_{\odot}$) and temperatures (40--90 K) of the clusters' local interstellar medium.  
Extinctions derived from the $JHK_S$, H$\alpha$, and UV images are similar to within a factor of 2 or 3.  These results suggest that eleven of the fourteen radio-selected objects are optically-visible young star clusters with a surrounding H~{\sc ii} region, that two are background objects, possibly AGN, and that one is a Wolf-Rayet star with a surrounding H~{\sc ii} region.  
\end{abstract}


\section{INTRODUCTION}

A galaxy's morphological evolution and the properties of its interstellar medium (ISM) are driven by massive stars in young (0-10 Myr) star clusters.  At these ages, star clusters produce metals from supernovae explosions, generate ionizing photons that determine the phase of the ISM, and provide mechanical energy to rearrange the ISM.  These effects originate in the local ISM but eventually contribute to galaxy-wide ISM changes \citep[see][]{calzetti04}.  Thus, the small-scale effects of young star clusters on their ISM have evolutionary consequences on the host galaxy in general.

When a young star cluster first forms, the OB stars in that cluster have not had time to remove their natal molecular cloud material, either by photoionization or by the mechanical force from the stellar winds.  Because this molecular material exhibits optical depths greater than 1 at visible wavelengths, the cluster should be detectable only at infrared (IR) and radio wavelengths.  \citet{johnson01} found 14 thermally-emitting objects in M 33 using high spatial resolution radio continuum maps at 6 cm and 20 cm.  These selected candidates have a spectral index ($\alpha^{6 cm}_{20 cm}$) of 0 or greater, indicating optically thick free-free emission from a thermal source.  The Lyman continuum photon rates determined from radio continuum maps imply the presence of 2-4 O7 equivalent stars and hence an age of 7 Myr or less.  

To learn more about these radio-selected objects, we undertook a multiwavelength survey from the radio to the far-ultraviolet (far-UV).  With the combination of {\it Spitzer} InfraRed Array Camera \citep[IRAC;][]{fazio04} and Multiband Imaging Photometer for {\it Spitzer} \citep[MIPS;][]{reike04} images, we identify the radio-selected objects from \citet{johnson01} and produce spectral energy distributions (SEDs) to determine their dust properties (temperature and mass) as well as morphologies of the polycyclic aromatic hydrocarbon (PAH)/dust emission.  With a near-IR color-color plot and absolute magnitudes, we determine their stellar ages, masses, and extinctions.  Finally, archival H$\alpha$ and ultraviolet (UV; GALEX) images were used to detect objects at shorter wavelengths for which we measured UV and H$\alpha$ extinctions based on stellar synthesis models.  The observations and data reduction are presented in \S \ref{obs}. Basic results of our analysis are given in \S \ref{res}, and discussion of our findings is given in \S \ref{dis}.  In \S \ref{summary} we summarize our findings.  

\section{OBSERVATIONS AND REDUCTIONS \label{obs}}

\subsection{{\it Spitzer} IRAC and MIPS Images \label{spitzerobs}}

{\it Spitzer} MIPS and IRAC images of M 33 were produced under the Guaranteed Time Observer (GTO) program ``M 33 Mapping and Spectroscopy" (PID 5) by {\it Spitzer} Science Working Group member R. D. Gehrz. \citet{hinz04} describe the MIPS observations used in this article.  Observations of M 33 were made using all four bands of IRAC on 2004 January 9.  The M33 IRAC mapping sequence consisted of 438 frames per channel, including a 3 point $1/2$ pixel dither for each map position. The integration time was 12 s per frame. The raw {\it Spitzer} IRAC data (single epoch) were processed and flux calibrated with version 10.0.1 of the {\it Spitzer} Science Center (SSC) pipeline.  Post-basic calibrated data (Post-BCD) processing was carried out using the 2004 June 6 Linux version of the SSC MOsaicking and Point source EXtraction (MOPEX) software \citep{makovoz05}. Four steps of MOPEX were implemented: cosmetic fix, background matching, outlier detection, and mosaicking.  The cosmetic fix module reduced column pull-down and mux-bleed (trails that form rightward of bright objects) artifacts associated with 3.6~\micron \ and 4.5~\micron \ data. Background matching was performed by minimizing the pixel differences in overlapping areas with respect to a constant offset computed by the program. Cosmic rays and other outliers were detected and eliminated with the outlier detection module. In the final step, the images were reinterpolated to a pixel scale of approximately 0\farcs86 pixel$^{-1}$.  The final science images are in units of MJy sr$^{-1}$.

\subsection{SQIID $JHK_S$ Images \label{nearirobs}}

The near-IR $JHK_S$ data were obtained 2002 October 16-23 on the 2.1 m telescope at Kitt Peak National Observatory using the NOAO SQIID near-IR instrument.  SQIID obtains simultaneous images in $JHK_S$ at a pixel scale of 0\farcs7 pixel$^{-1}$ over a 304 \arcsec~$\times$ 317 \arcsec~area.  One night was photometric while another night was sporadically photometric.  We observed each radio-selected object position with a standard 5 s integration time, 24 coadds, and 3 sets of 5 dithered images (total integration times of 75 s).  Throughout these nights, we observed 3-6 UKIRT faint standard stars \citep{hawarden01} for flux calibration.  

All data reductions were performed using the {\it Image Reduction and Analysis Facility}\footnote{IRAF is distributed by the National Optical Astronomy Observatories, which are operated by the Association of Universities for Research in Astronomy, Inc., under cooperative agreement with the National Science Foundation.} (IRAF) software. Averaged dark images were subtracted from individual images, followed by flatfielding with the averaged flatfield image for the appropriate filter.  We created sky images from each set of dithered science images and then subtracted the scaled sky background images from each science image in their respective sets.  The science images were then registered and combined.  Finally, a world coordinate system was generated for each $JHK_S$ image from field stars' celestial coordinates in the 2MASS point source catalog.  

Flux calibration was performed with the standard stars.  Zero points were determined using a 7\arcsec~aperture; we found that color terms were negligible.  Between the flux calibration of the photometric and near-photometric nights, the zero point differences were 0.01 magnitudes for J and H, and 0.08 magnitudes for K$_{\rm S}$.  We compared our derived magnitudes of field stars to those found in 2MASS, appropriately transformed to the UKIRT system \citep{cutri03}.  These magnitudes were found to be in close agreement ($\sim$0.05 mag for all filters).

\subsection{Archival Imagery \label{archivedata}}

We searched multiwavelength data archives for images of these radio objects.  We found four other wavelength bands that had suitable data.  VLA 6 cm continuum maps of the fourteen radio-selected objects were generated from the data taken for  \citet{duric93}.  These radio maps were used to discern which object in the {\it Spitzer} images is the actual compact radio object, and these radio maps were also used by \citet{johnson01}.  The SQIID near-IR observations did not cover all M 33 radio-selected objects.  Therefore, we used the M 33 2MASS mosaic produced by \citet{jarrett03}.  This image allowed us to derive $JHK_S$ magnitudes and colors for \citet{johnson01} objects 1--4 and 12.  R-band and H$\alpha$ images of M 33 were taken from \citet{massey02}.  Continuum emission in the H$\alpha$ image was removed by subtracting a scaled R-band image.  We determined the scaling factor by comparing the aperture fluxes of 20 stars in both images. 
To determine what the corresponding flux is for one count (data number or  DN) in the continuum-subtracted H$\alpha$ image, we used H$\alpha$ fluxes of three round, clearly separated H~{\sc ii} regions given in \citet[][; BCLMP\footnote{\citet{boulesteix74}} 71a, 95Ab, and 633a$+$b]{hodge02}.  \citet{hodge02} give H$\alpha$ fluxes as well as the areas for the H~{\sc ii} regions.  We choose isolated, spherical H~{\sc ii} regions in order to both avoid contamination from other sources and to make use of simple circular aperture photometry.  The conversion factor derived is 7.64$\times$10$^{-19}$ ergs s$^{-1}$ cm$^{-2}$ DN$^{-1}$.  
Far-UV (0.1528 $\mu$m) and near-ultraviolet (near-UV; 0.2271 $\mu$m) archival images of M 33 were downloaded from the GALEX data archive to be used for aperture photometry.  Flux calibration of the GALEX data from DN s$^{-1}$ to Jy is given in the GALEX data handbook\footnote{http://galexgi.gsfc.nasa.gov/Documents/ERO$\_$data$\_$description$\_$2.html}.

\section{RESULTS \label{res}}

\subsection{Finding the Radio-Selected Objects \label{find}}

We located the 14 objects from \citet{johnson01} in the far-UV (0.1528 $\mu$m), near-UV (0.2271 $\mu$m), H$\alpha$, $JHK_S$, and {\it Spitzer} images.  Figures \ref{m33object1} through \ref{m33object14} show 90\arcsec~sqaure images centered on each radio position.  Because the far-UV and near-UV images are similar, only the near-UV is used in Figures \ref{m33object1}--\ref{m33object14}.  All images have an orientation of north up, east left.  In all figures we have placed a $\sim$9\arcsec~circle around the center coordinate from \citet{johnson01} to reflect the positional uncertainties, which are much larger than the positional uncertainties of all other images in this study.  For all images, the wavelengths are designated in the upper right corner.  The upper left corners of the near-UV and 8 $\mu$m images show the object's designation (e.g., 1 for Figure \ref{m33object1}).  The $J$ and 3.6 $\mu$m images give the labeling of the objects that we think are the radio-selected sources (discussed below).  In the 8 $\mu$m images we have encircled any extended dust/PAH shells near the radio position.  In the 4.5 $\mu$m image, we have superimposed contours of our  6 cm radio maps.  

We identify potential IR counterparts to the radio emission in the {\it Spitzer} images.  {\it Spitzer} images are the obvious choice because they are the closest in wavelength to the radio maps.  Also, if these objects are embedded sources, then the dust emission around these objects should be very strong and should obscure emission from shorter wavelengths.  Below, we discuss the object determination on a case-by-case basis.  The celestial positions of these objects described below are given in Table \ref{tab1}.  

\notetoeditor{please place the figures on a two page spread like opening a book.  The left side should be the UV through 5.8 micron images.  The right should be the 8 through 160 micron images with the caption below it.   Text can be used to fill in any gaps on the right or left-hand side.  This easy comparison of all the images will aid the reader in doing their own comparison for a single object. } 
\placefigure{m33object1}
\placefigure{m33object2}
\placefigure{m33object3}
\placefigure{m33object4}
\placefigure{m33object5}
\placefigure{m33object6}
\placefigure{m33object7}
\placefigure{m33object8}
\placefigure{m33object9}
\placefigure{m33object10}
\placefigure{m33object11}
\placefigure{m33object12}
\placefigure{m33object13}
\placefigure{m33object14}

\noindent {\bf Object 1}: At all wavelengths in Figure \ref{m33object1}, excluding UV, the brightest potential source is found at the northeast position of the uncertainty circle around the object's position.  However, the 6 cm emission peaks to the southwest of this object.  A source appears coincident with this radio emission in the near-UV, 3.6 $\mu$m, and 4.5 $\mu$m images, and we designate this as object 1.  Photometry of this radio-coincident source is contaminated by the closer, brighter source.  The lack of detection at 24 $\mu$m of object 1 is probably inconsistent with this object being an example of a young, embedded phase of star cluster formation.  Object 1 is possibly a background object, and therefore, we have eliminated it from further discussion.  \citet{johnson01} state that some of their sample could be background galaxies.  

\noindent {\bf Object 2}:  The radio position from \citet{johnson01} is located inside a Wolf-Rayet nebula, termed BCLMP 638 or IC 132.  The Wolf-Rayet star ([MC83]\footnote{\citet{massey83}} 23, a WN star) is located in the southwest part of the uncertainty circle in both the 3.6 $\mu$m and the  B image in \citet{johnson01}.  We designate object 2 as the prominent dust/PAH emission part of BCLMP 638 located at the southeast portion of the uncertainty circle.  The MIPS 24 $\mu$m emission is bright in this region as well.  The other MIPS bandpasses cannot identify whether the dust emission arises around the Wolf-Rayet star or object 2 because of lower resolution and larger pixel size of the detectors.  The 6 cm radio contours suggest that the bright radio emission originates around the Wolf-Rayet star and two bright sources, probably stars, 9\arcsec~west of the Wolf-Rayet star.  The H$\alpha$ image indicates that IC 132 is a bipolar nebula.  Knowing the size and shape of the nebula, we suspect that the bright 6 cm knots of emission are found coincidentally on top of the two bright sources in the 4.5 $\mu$m image and that the bright 6 cm knots are actually bright radio knots along the edge of the bipolar nebula.  At $JHK_S$ wavelengths, we see the Wolf-Rayet star, but no emission is found at the bright dust/PAH emission source to the east of the Wolf-Rayet star.  At UV wavelengths, the Wolf-Rayet star shows up prominently in the images, but object 2 is not detected at all.

\noindent {\bf Object 3}:  We identify the IRAC object in the middle of the uncertainty circle with the radio-selected source.  The 6 cm radio contours and $JHK_S$ emission are coincident with this IRAC source.  Some H$\alpha$  emission is present for this object.  Note that the dim IRAC object to the southwest of object 3 is just as bright as object 3 at H$\alpha$.  \citet{johnson01} also detect this object in their B image, and their radio position is coincident with the H~{\sc ii} region IC 133 or BCLMP 623.  Finally, object 3 is also detected at UV wavelengths, implying little extinction for this object.  

\noindent {\bf Object 4}:  Like object 1, this object is detected only at IRAC wavelengths and is located at the center of the radio uncertainty circle.  The 6 cm contours in the 4.5 $\mu$m image show that this object is coincident with the radio emission.  The lack of detection at 24 $\mu$m of object 4 is probably inconsistent with this object being an example of a young, embedded phase of star cluster formation.  Object 4 is possibly a background object, and therefore, we have eliminated it from further discussion.  \citet{johnson01} state that some of their sample could be background galaxies.

\noindent {\bf Object 5}:  The celestial position of object 5 is coincident with the H~{\sc ii} regions BCLMP 35/36 \citep{johnson01}.  Two objects are found at the outer edge of the uncertainty circle for the radio position.  We term the north object ``a" and the southern object as ``b."  Looking at the MIPS images, object 5a is the dominant 24 $\mu$m source, but object 5b is also present but dimmer. The 6 cm contours show that object 5a is the brighter radio object but that object 5b has some associated radio emission.  The two objects appear to be enveloped by a diffuse region of 6 cm emission. Because of this 6 cm connection, we have labeled a third object (``5s") as the combination of the two.  Object 5s is designated in the 8 $\mu$m image.  In the $JHK_S$ images, we do not detect either of the objects.  At H$\alpha$, a complex ionized gas morphology surrounds these objects.  Object 5a is detected at H$\alpha$ and the morphology around this point appears shell-like.  Object 5b is the brightest H$\alpha$ part of a shell-like complex south of the uncertainty circle.  The UV and B \citep[see][]{johnson01} images show a point source in the center of the uncertainty circle that is not associated with objects 5a and 5b.  This UV-optical point source is close to the center of object 5s.  

\noindent {\bf Object 6}:  In the 3.6 $\mu$m image of Figure \ref{m33object6}, a star-like source is found at the southeastern portion of the uncertainty circle, and we term this object ``6a".  Northeast of the uncertainty circle is a source of bright PAH and dust emission, which we term ``6b".  In the MIPS 24 $\mu$m image, these sources are enveloped by diffuse emission and may be interrelated.  In the other MIPS images, the dominant dust emission appears to be from object 6a.  In the 8 $\mu$m and 24 $\mu$m images, a shell of PAH emission is close to the radio position, which we label as object ``6s."  Faint 6 cm emission surrounds objects 6a and 6b, but no compact emission is associated with either IR source.  In the $JHK_S$ images, we can detect object 6a only.  The H$\alpha$ morphology is quite complex with several shells around the radio position.  Object 6a does not appear to have any H$\alpha$ emission, while a few points of H$\alpha$ emission coincident with object 6b.  H$\alpha$ emission appears to fill the cavity of object 6s.  Object 6a is detected faintly in the B image of \citet{johnson01}, and the celestial position is coincident with the H~{\sc ii} regions BCLMP 39/40.  No UV emission is apparent for any of the objects.  

\noindent {\bf Object 7}:  This radio position is coincident with [GDK99] 91, and \citet{gordon99} claim that this object is a H~{\sc ii} region.  An IRAC and MIPS point source (object 7a) is found almost exactly at the radio position.  No 6 cm compact radio emission from our maps is associated with this object.  The object ``7b" is found well north of the radio position and is a compact 6 cm source.  At 70 $\mu$m object 7b dominates the emission.  Perhaps  object 7b is the compact radio source 7 from \citet{johnson01}.  Both objects are present in the $JHK_S$ images and at H$\alpha$.  The gas morphologies around those objects appear to be simply spherical for both objects.  In the GALEX UV images, only object 7b is detected.  
  
\noindent {\bf Object 8}:  In the IRAC images, a bright source appears in the southeastern part of the uncertainty circle.  No additional sources appear around the radio position in either MIPS or IRAC images.  This bright IRAC source is also the source of the 6 cm emission.  At 8 $\mu$m the source appears to be associated with a long extended line of PAH/dust emission.  However, at H$\alpha$, this arc of emission disappears, and a simple sphere explains the H$\alpha$ morphology.  This source is also detected in the B image in \citet{johnson01}, and the radio position is coincident with the H~{\sc ii} region BCLMP 703.  Some UV emission is also apparent for this object. We did not detect object 8 in the $JHK_S$ images.  

\noindent {\bf Object 9}:  A bright point source appears in the uncertainty circle at all {\it Spitzer} wavelengths and in the JHK$_S$ images.  The 6 cm radio emission is also centered on this {\it Spitzer} source.  A bright H$\alpha$ source is found on top of the location of object 9.  An arc of H$\alpha$ emission is present to the north of object 9. This object is also detected in the B image of \citet{johnson01}, and the radio position is coincident with the H~{\sc ii} region BCLMP 79.  No UV emission is apparent for object 9.  

\noindent {\bf Object 10}: Two point sources appear at opposite edges of the uncertainty circle at 3.6 $\mu$m in Figure \ref{m33object10}.  We have termed the eastern most ``10a" and the other ``10b."  Object 10a dominates at 24 $\mu$m, while object 10b dominates the PAH emission at 8 $\mu$m.  The 6 cm contours envelope  both objects with some radio emission connecting the two points, similar to object 5.  
%suggesting that the object 10s is the best choice.  
Thus, we have also labeled a set of flux measurements with ``10s" for the combination of the two.  
No sources are detected in the $JHK_S$ images.  At H$\alpha$, both sources are on the edge of a shell-like complex of H$\alpha$ emission.  The center of this shell is devoid of H$\alpha$, but according to the 6 cm contours, some thermal radio emission is apparent.  At H$\alpha$, these objects do not look interconnected as when looking at longer wavelengths.  Object 10a is detected in the B image in \citet{johnson01}, and the radio position is coincident with the H~{\sc ii} region BCLMP 87.  Near-UV and far-UV emission emanates from object 10a only but not from object 10b.  

\noindent {\bf Object 11}:  The only bright star-like source near the radio position is found in the northeastern part of the uncertainty circle in the 3.6 $\mu$m image in Figure \ref{m33object11}.  We have labeled this source as ``11a."  However, the 6 cm contours of the brightest radio emission are not coincident with this object but are associated with a ring of dust/PAH emission found around the radio position uncertainty circle.  This shell is labeled object 11s in the 8 $\mu$m image. The bright dust/PAH sources in this circle are coincident with the 6 cm emission points.  
However, the B image of \citet{johnson01} shows a clear source in the middle of the uncertainty circle that we see only at UV.  This source is not object 11a because we also detect object 11a in the UV image.  The radio position is coincident with the H~{\sc ii} region BCLMP 77.  

\noindent {\bf Object 12}: The only source near the radio position is found in the southwestern part of the uncertainty circle in the 3.6 $\mu$m image of Figure \ref{m33object12}.  In the MIPS images this source appears to be the dominant dust emitter.  The 6 cm contours also lie directly on top of object 12.  Object 12 was not detected in the $JHK_S$ images.  At H$\alpha$ object 12 appears to have a spherical morphology with no shell-like emission around the object.  This object is also detected in the B image of \citet{johnson01}, and the radio position is coincident with the H~{\sc ii} region BCLMP 714.  Object 12 is also detected at UV wavelengths.  

\noindent {\bf Object 13}: In Figure \ref{m33object13} bright 3.6 $\mu$m source is found in the middle of the uncertainty circle.  In the MIPS and IRAC images, no other dominant source exists around the radio position.  The 6 cm contours are also coincident with the IR source.  The H$\alpha$ emission appears spherical for object 13.  A crescent-shaped object is located inside the uncertainty circle in the northeastern portion of the B image in \citet{johnson01}, and the radio position is coincident with the H~{\sc ii} region BCLMP 712.  Some UV emission is found to the south of the uncertainty circle but is not associated with object 13.  

\noindent {\bf Object 14}:  A star-like source appears in the southwestern part of the uncertainty circle (see the 3.6 $\mu$m image of Figure \ref{m33object14}).  We have termed this object ``14a."  The dust and PAH emission in MIPS and IRAC images appear to be in a cometary V-shape to the south of the radio position.  Assuming that this V-shaped dust/PAH emission is related, we have labeled it ``14s" and provided a circle for reference in the 8 $\mu$m image.  The 6 cm contours in the 4.5 $\mu$m image show no significant radio emission around this object.  Some faint emission is associated with object 14a,  but no 6 cm emission appears to be associated with object 14s.  At H$\alpha$ the cometary emission of object 14s at 8 $\mu$m is only a small portion of the shell of H$\alpha$ emission around object 14a.  This shell of emission is oblong with the major axis along the northeast.  No H$\alpha$ emission is associated with object 14a.  Object 14a is also detected in the B image of \citet{johnson01}, and the radio position is coincident with the H~{\sc ii} regions BCLMP 749/750.  Object 14a is detected at UV wavelengths.

\subsection{Aperture Photometry of the Radio-Selected Objects in M 33}

The aperture used in the near-IR (SQIID and 2MASS), IRAC, H$\alpha$, and GALEX UV images for the star-like objects (i.e., those that are not shell apertures) had a radius of 7\arcsec.  The inner and outer radii of the sky annulus for background subtraction were 7\arcsec~and 14 \arcsec~respectively, making the sky area greater than that of the aperture.  For object 2, a significant bipolar nebula is present in H$\alpha$ that is not seen with the {\it Spitzer} data.  For deriving a H$\alpha$ flux, we extended the sky annulus well outside this large nebula to remove only sky emission.  The shell apertures (i.e., those termed with a ``s") along with their sky annulus are much larger and set on an individual basis to encompass the brightest emission from these shells.  Aperture photometry for the ``s" designated objects was not done with the GALEX UV data, but was done for the H$\alpha$ images since the shells were detected at H$\alpha$.  The $JHK_S$ magnitudes and colors as well as H$\alpha$ and UV fluxes are presented in Table \ref{tab1}.  These magnitudes, colors, and fluxes are not corrected for Galactic or internal reddening.  For those sources not detected in any of the $JHK_S$, H$\alpha$, or UV images, we give 3$\sigma$ upper limits for the fluxes and lower limits for magnitudes.  

Aperture corrections were mandatory for all IRAC fluxes.  We generated our own aperture corrections for our specific aperture sizes using the theoretical point spread function (PSF) images provided by the SSC\footnote{2004 January In-Flight PSF images downloaded from http://ssc.spitzer.caltech.edu/irac/psf.html}.  Since the aperture and sky annulus were constant for all star-like IRAC photometry, the multiplicative values are 1.135, 1.14, 1.176, and 1.285 for IRAC 3.6 $\mu$m, 4.5 $\mu$m, 5.8 $\mu$m, and 8 $\mu$m respectively.  Aperture corrections for the shell photometry were also generated with the theoretical PSF images and are $\sim$1.  See Table \ref{tab2} for the IRAC fluxes.

For the MIPS images, we used a constant aperture of 48\arcsec~for all 3 channels.  This aperture size is the smallest reasonable aperture size for the 160 $\mu$m images where the FWHM is $\sim$40\arcsec.  Sky subtraction was also performed with the area of the sky annulus greater than or equal to the area of the source aperture.  The MIPS science images were in the units of DN s$^{-1}$.  The conversion factors from DN s$^{-1}$ to $\mu$Jy arcsec$^{-2}$ were 1.032, 14,900, and 1,000 for the 24 $\mu$m, 70 $\mu$m, and 160 $\mu$m images respectively.  Aperture corrections were also mandatory for all MIPS fluxes.  We generated the aperture corrections from theoretical PSF images\footnote{downloaded PSF images from http://ssc.spitzer.caltech.edu/mips/psffits/ on 2004 April} using the same size apertures and sky subtraction areas as those used for the MIPS photometry.  These aperture corrections were 1.13, 1.61, and 1.89 for the 24 $\mu$m, 70 $\mu$m, and 160 $\mu$m images respectively.  The MIPS fluxes for the objects are given in Table \ref{tab2}.  Because the resolution of the MIPS images is lower than those of IRAC and SQIID/2MASS, the MIPS fluxes are given in the shell designation for those objects with shells (e.g., 11s) or the best possible candidate (e.g., object 3).

\subsubsection{Uncertainties}

Uncertainties for all fluxes and magnitudes are given.  For the IRAC fluxes, the uncertainties account for Poisson noise, variation in the background, correction for correlation in the subsample images, and $\pm6$\% 
uncertainty caused by flux calibration.  For the MIPS fluxes, the uncertainties accounted for are similar to those for IRAC, but the flux calibration uncertainty is $\pm10$\% 
for MIPS 24 $\mu$m and $\pm$20\% 
for MIPS 70 $\mu$m and 160 $\mu$m.  
For the $JHK_S$ magnitudes, H$\alpha$ fluxes, and UV fluxes, we account for Poisson noise and variation in the background.  These uncertainties are presented in Tables \ref{tab1} and \ref{tab2}.

\subsection{Extinctions, Ages, and Masses Derived From the $JHK_S$ Images \label{eamjhk}}

In Figure \ref{m33jhk} we plot the colors of the nine radio-selected objects with $JHK_S$ detections in all three bandpasses.  Also plotted are the $J$--$H$ and $H$--$K_S$ colors as derived from the Starburst 99 v4.0 SEDs \citep{leitherer99}.  The important parameters for these instantaneous burst models are fixed stellar masses of 10$^6$ $M_{\odot}$, upper and lower stellar mass limits of 100 $M_{\odot}$ and 1 $M_{\odot}$, and a Salpeter initial mass function (IMF).  The solid line shows model colors for a metallicity of Z = 0.008. The dashed line shows model colors for a metallicity of Z = 0.02.  The metallicity of each object was estimated using the M 33 oxygen abundance gradient equation from \citet{garnett97}.  To derive the metallicities, we assumed that the objects' metallicities relative to solar were proportional to the ratio of their oxygen abundances.  Since Starburst 99 models have only a few discrete metallicities, we used the closest Starburst 99 metallicity, typically Z = 0.008 or Z = 0.02.  

We do not use the colors for object 2 to derive extinctions, ages, and masses because the colors found for object 2 are for the Wolf-Rayet star.  Thus, the central photoionizing source is not a star cluster. We could not determine colors for objects 1, 4, 5, 8, 12, and 13 because they are not detected in all three near-IR bands.

\subsubsection{Extinctions \label{extinct}}

Extinction measurements were estimated for seven objects by extrapolating the measured colors back along the reddening vector to the unreddened Starburst 99 model colors.  We know the cluster's color, the slope of the extinction vector \citep{schlegel98}, and the theoretical colors.  The distance between the cluster color and unattenuated model color was converted into a $K_S$ extinction, $A_{K_S}$ \citep{schlegel98}.  The results of these extinction derivations are found in Table \ref{tab3}, and the values range from $A_{K_S}$ of 0.23 mag to 1.31 mag.  These extinctions, as well as all other derived extinctions, have not had Galactic reddening removed.  The K$_S$ extinction caused by the Galaxy is 0.015 mag approximately \citep{schlegel98}.  Thus, the actual range of reddening caused by the star clusters and M 33 is 0.21--1.29 mag.

\subsubsection{Ages} 

All points on the Starburst 99 model track have an associated age.  Moving the observed cluster colors back along the reddening vector, the observed colors yield an estimate of the cluster age.  The 1$\sigma$ age uncertainties are based from the 1$\sigma$ color uncertainties. The cluster ages and their uncertainties are given in Table \ref{tab3} and range from 2 Myr to 10 Myr (i.e., young clusters).  

We could not derive an age for cluster 9 because the cluster shows a significant $H$--$K_S$ excess indicative of dust emission and/or emission lines.  These excesses have been seen in star clusters with ages less than 6 Myr \citep{buckalew05}.  Thus, by comparison to previous results, we conclude that the $H$--$K_S$ excess implies an age for cluster 9 of less than 6 Myr.  

\subsubsection{Masses}

Once the extinctions and ages are known, we can use a similar procedure to the one outlined in \citet{buckalew05} to derive the masses. Instead of using the blue absolute magnitudes, we use unreddened absolute $JHK_S$ magnitudes to derive the stellar masses.  
We compared our derived absolute magntiudes with those derived from the 10$^6$ $M_{\odot}$ Starburst 99 model at the appropriate age and metallicity.  
The scaling factor between the model's absolute magnitude and the one that we derived gave us the scaling factor to apply to the mass.  Multiplying the scaling factor to the model mass gives us the derived stellar mass for the cluster.   
Table \ref{tab3} lists these masses, which have a range of 10$^3$--10$^{3.5}$ $M_{\odot}$.

\subsection{Spectral Energy Distributions \label{models}}

The UV to mid-IR SEDs are plotted in Figures \ref{m33sed1}--\ref{m33sed3}.  The size of the symbols used in the SEDs are equivalent to the flux uncertainties. Upper limits are given as downward-pointing arrows.  
Using these SEDs, we model the free-free and dust emission around these objects.

\subsubsection{Dust emission \label{dustem}}

Using the 3 MIPS flux results, we derive dust masses and temperatures.  We derive the dust temperature using simple blackbody fits of the 3 MIPS fluxes.  With only 3 data points, amplitude and dust temperature are the free parameters.  In some cases, the data were best fit by two blackbodies with the same amplitude but different temperatures.  The best fit was found for the set of parameters that minimized the residuals between the model and measured fluxes.  Two blackbodies were only used when such a fit produced a factor of 10 decrease in the residuals (objects 2, 3, and 10).  The best-fit temperatures are found in Table \ref{tab3} and range between 60 K and 90 K.  

Dust masses were calculated using the procedure outlined in \citet{dale01}.  If only one temperature was used to model the dust emission, the MIPS 70 $\mu$m flux was used to calculate the dust mass.  If two blackbodies provided a  better fit, the MIPS 70 $\mu$m and 160 $\mu$m fluxes were used to derive the dust masses with the 70 $\mu$m flux used for the hotter of the two temperatures.  The dust masses are given in Table \ref{tab3} and range between 1 and 1000 $M_{\odot}$.

\subsection{Extinctions from H$\alpha$ and UV Fluxes \label{extinct2}}

Using the H$\alpha$ fluxes from Table \ref{tab1} and radio fluxes from \citet{johnson01}, we calculate H$\alpha$ extinctions using the equations detailed in \citet{caplan86}.  We assume a temperature of the H~{\sc ii} region of 10,000 K, a N(He$^+$)/N(H$^+$) of 0.1, densities of 100 cm$^{-3}$, and 100\% 
thermal emission at 6 and 20 cm. We compute extinctions for H$\alpha$ using both the 20 cm and 6 cm fluxes from \citet{johnson01}.  We present the average H$\alpha$ extinctions ($A_{{\rm H}\alpha}$) in Table \ref{tab3}.  The extinctions range from 0.72--3.46 mag.  Extrapolating these extinctions to $A_{K_S}$ equivalents using the extinction law from \citet{schlegel98} and removing Galactic reddening, the range of $A_{K_S}$ extinctions is 0.081--0.44 mag.  

We computed UV extinctions by comparing the measured UV emission of the objects to GALEX FUV and NUV fluxes calculated from Starburst 99 SEDs.  The important parameters (metallicity, stellar mass, and age) for each SED came from the results derived with the $JHK_S$ colors and magnitudes. The other standard parameters of Starburst 99 were assumed (Salpeter IMF, etc.) as they were detailed in \S \ref{eamjhk}.  The GALEX system response curves were taken from the GALEX data handbook.  The fluxes generated with the filter response curves and Starburst 99 SEDs were assumed to be the unreddened FUV and NUV fluxes.  Ratioing the measured and model fluxes, we calculate extinctions for 0.2271 $\mu$m and 0.1528 $\mu$m ($A_{0.2271}$ and $A_{0.1528}$) and present these extinctions in Table \ref{tab3}.  Three objects had the necessary UV fluxes and $JHK_S$ ages and masses to derive these extinctions.  The values of $A_{0.2271}$ are 3.88 mag, 0.686 mag, and 0.15 mag, and the values of $A_{0.1528}$ are 4.2 mag, 1.9 mag, and 0.43 mag for objects 3, 11a, and 14a respectively.  Converting these to equivalent $A_{K_S}$ extinctions using a factor of 15 (based on the $A_U/A_K$ from Schlegel et al.\ 1998) and removing Galactic reddening, the A$_{K_S}$ range is 0--0.26 mag.

\section{DISCUSSION \label{dis}}

\subsection{Morphologies of the Radio-Selected Objects \label{dismorphology}}

These objects in M 33 exhibit a diversity of PAH/dust morphologies in the IRAC 5.8 $\mu$m and 8 $\mu$m images and the MIPS 24 $\mu$m image.  We combine these IR morphologies with the morphologies seen at H$\alpha$ to determine a multiwavelength classification of the morphologies.  We find that a simple classification scheme of shell and spherical  morphologies applies best to these objects.   Objects 3, 7, 8, 9, 12, and 13 are classified as spherical, and objects 2, 5, 6, 10, 11, and 14 are classified as shells. We present an IRAC color-color diagram in Figure \ref{m33iraccolor} using the 8 $\mu$m, 5.8 $\mu$m, and 4.5 $\mu$m fluxes from Table \ref{tab2} for all objects.  The MIPS fluxes from Table \ref{tab2} are plotted in the color-color diagram in Figure \ref{m33mipscolor}.  We conclude that no clear separation of the different morphologies can be discerned from either color-color figure.

\subsection{Individual Descriptions of the Objects \label{inddis}}

\noindent {\bf Object 2}:  A bipolar nebula around a Wolf-Rayet star makes up this thermal radio source.  We find an H$\alpha$ extinction ($A_{{\rm H}\alpha}$) of 0.72 mag.  Dust masses and temperatures were 12 $M_{\odot}$ at 64 K and 330 $M_{\odot}$ at 35 K.  The 12 $M_{\odot}$ of dust at 64 K could be ejected stellar material from the WN Wolf-Rayet star.  The colder dust mass is more likely dust from its natal molecular cloud and/or swept-up ISM.  

\noindent {\bf Object 3}:  Object 3 (IC 133) is one of the brightest point sources in M 33 at 24 $\mu$m.  This object appears to have a spherical morphology of ionized gas (H$\alpha$) and dust (8 $\mu$m, 24 $\mu$m) around the central star cluster.   Fitting two blackbodies, we derive dust temperatures of 87 K and 44 K and dust masses of 31 $M_{\odot}$ and 810 $M_{\odot}$ respectively.  The age of this star cluster is 7 Myr, and the stellar mass is 10$^{3.9}$ $M_{\odot}$.  The extinction derived from the $JHK_S$ color-color diagram is a factor of 1.7 larger than that derived from H$\alpha$ and the UV. 

\noindent {\bf Object 5}:  This multiple-peaked IRAC source is classified as a shell morphology in our simplified scheme.  From the MIPS data, we derive a dust temperature of 66 K and a mass of 12 $M_{\odot}$.  The H$\alpha$ extinction for the shell is 0.51 mag.  

\noindent {\bf Object 6}:  Object 6 is interesting for the shell structure seen at 8 $\mu$m that lies outside the uncertainty circle  of the radio position.  Object 6a is a stellar source inside the uncertainty circle but is found at the northwestern edge of this shell.  Object 6b appears as a bright 8 $\mu$m source on the edge of this shell and is also apparent at 3.6 $\mu$m.  Object 6a is one of the youngest star clusters in this sample with an age of 3 Myr, and its derived stellar mass is 10$^{3.3}$ $M_{\odot}$.  The dust mass derived for this system is 22 $M_{\odot}$ at a temperature of 61 K.  The derived extinctions from the $JHK_S$ and H$\alpha$ images are similar ($A_{K_S}$ extinctions of 0.23 mag and equivalent $A_{K_S}$ of 0.24 mag respectively).

\noindent {\bf Object 7}:  Object 7a is a point source found at the exact location of the radio position.  However, object 7b is also a bright point source and has bright 6 cm contours on top of its position (something not seen with object 7a; see Figure \ref{m33object7}).  We find that object 7a has a very young age of 2 Myr and a stellar mass of 10$^{3.7}$ $M_{\odot}$. Object 7b is older at an age of 10 Myr and a stellar mass of 10$^4$ $M_{\odot}$.  The dust characteristics, which we derive for the synthesis of the two dominant objects, are a dust temperature of 73 K and mass of 7.6 $M_{\odot}$.  Object 7a is detected at H$\alpha$ but not UV.  Object 7b is detected at both wavelengths.  The $JHK_S$ extinction for object 7a is 1.4 times higher than that derived from H$\alpha$ ($A_{K_S}$ extinctions of 0.54 mag and equivalent $A_{K_S}$ of 0.14 mag respectively).  The $JHK_S$ extinction for object 7b is 3 times higher than the extinction derived from H$\alpha$ ($A_{K_S}$ of 1.31 mag and equivalent $A_{K_S}$ of 0.09 mag respectively).  The UV extinction derived for object 7b is 2.5 times smaller than that from the $JHK_S$ images ($A_{K_S}$ of 0.31 mag and equivalent $A_{K_S}$ of 1.31 mag respectively).  

\noindent {\bf Object 8}: 
Object 8 is a bright ellipsoid found at the southeastern section of the uncertainty circle (see Figure \ref{m33object8}).  The point source is found along a line of unassociated extended 8 $\mu$m emission.  Because the object was not detected in the $JHK_S$ images, no stellar mass, age, or extinctions were derived.  We derived an A$_{{\rm H}\alpha}$ of 0.36 mag.  The dust characteristics are typical with a temperature of 66 K and a mass of 16 $M_{\odot}$. 

\noindent {\bf Object 9}:  This object is a strong, compact source at all infrared wavelengths and is coincident with the 6 cm radio contours (see Figure \ref{m33object9}).  The $JHK_S$ color-color diagram in Figure \ref{m33jhk} shows that this cluster candidate has a significant $H$--$K_S$ excess.  The unusually red H$-$K$_{\rm S}$ of 1.1 mag may be caused by emission from hot dust or nebular lines in the K-band \citep[see][]{buckalew05} and implies an age less than 6 Myr.  The $J$--$H$ of 0.8 mag is attributable to extinctions of at least $A_{K_S}\sim0.68$ mag. The $JHK_S$ extinction is 1.2 times higher than the $A_{K_S}$ equivalent extinction derived from H$\alpha$.  Assuming an age of 6 Myr, we find that the stellar mass is 10$^{4.4}$ $M_{\odot}$, the highest in the sample. The dust characteristics of this object are odd with a dust temperature of 95 K, the hottest in the sample, and mass of 1.5 $M_{\odot}$, the smallest in the sample.

\noindent {\bf Object 10}:  Object 10 is classified as having a shell-like morphology because of its multiple-peaked but associated dust/PAH sources, similar to object 5.  Because objects 10a, 10b, and 10s were not detected in the $JHK_S$ images, no stellar mass, age, or extinctions were derived.  The extinctions at H$\alpha$ were the only derived and had the values of 2.45, 3.46, and 0.96 mag for objects 10a, 10b, and 10s respectively.  The dust characteristics of this system were best fit with two blackbodies with temperatures of 65 K and 40 K and with masses of 55 $M_{\odot}$ and 980 $M_{\odot}$. 

\noindent {\bf Object 11}:  Object 11a is a stellar source inside a shell of dust/PAH emission termed object 11s.  This shell is also the object that is associated with the 6 cm emission (see Figure \ref{m33object11}).  The $A_{K_S}$ of 0.79 mag from the $JHK_S$ images is 1.7 times larger than that derived from H$\alpha$ (equivalent $A_{K_S}$ of 0.22 mag).  The UV extinctions are 2 times smaller than that derived from the $JHK_S$ images.  The star cluster has an age of 7 Myr and a mass of 10$^{3.2}$ $M_{\odot}$.  The dust characteristics are a temperature of 57 K and a mass of 64 $M_{\odot}$.  The overall morphology of this object suggests the presence of a UV-bright star cluster at the center of a wind-blown shell, illuminating the surrounding ISM which emits strongly at mid and far-IR wavelengths.  Based on the age and morphology, this star cluster has similar results to old (an age greater than 6 Myr) star clusters from \citet{buckalew05}.  

\noindent {\bf Object 12}:  Object 12 is a spherical source detected from the UV to mid-IR.  We could not derive the stellar properties.  The extinction from H$\alpha$ implies a $A_{K_S}$ of 0.21 mag, not a large extinction for an embedded source.  The dust properties of this source are 15 $M_{\odot}$ at 61 K.

\noindent {\bf Object 13}:  Object 13 is not seen at near-IR nor UV but appears as a point-like source at MIPS, IRAC, and H$\alpha$.  The H$\alpha$ extinction of object 13 is $A_{{\rm H}\alpha}$ = 1.44 mag.  The dust properties of object 13 are a temperature 58 K and a mass of 23 $M_{\odot}$.

\noindent {\bf Object 14}:  A star-like source (object 14a) appears just outside the southwestern part of the uncertainty circle in Figure \ref{m33object14}.  Object 14a is surrounded by a cometary PAH/dust emission and a shell of H$\alpha$ emission.  The age of object 14a is 7.5 Myr, and the stellar mass derived is 10$^{3.4}$ $M_{\odot}$. Such an old star cluster should have a more evolved (i.e., shell-like) morphology \citep[see][]{buckalew05}.  The extinction from $JHK_S$ is 1.06 times larger than the H$\alpha$ extinction for object 14a.  The UV extinction of object 14a is 1.2 times lower than that from the $JHK_S$ images.  The dust properties of object 14 are a temperature of 56 K and a mass of 27 $M_{\odot}$.

\section{SUMMARY \label{summary}}

These radio-selected objects were thought to be embedded young star clusters, possibly larger analogues to ultracompact H~{\sc ii} regions.  We have observed the positions of these radio-selected objects in M 33 from the UV to the mid-IR.  We have detected the majority of the radio-detected candidates at all of these wavelengths.  These 11 radio-detected star clusters (excluding objects 1, 2, and 4) are young, when O stars are still present, but appear to have ages and extinctions that place them in the detectable limits of optical images.  All 11 star clusters are known optical H~{\sc ii} regions.  

Using the near-IR data, we derived ages, extinctions, and masses.  For the limited age results, they range from 2 Myr to 10 Myr.  The stellar masses derived from the near-IR data are typically 10$^3$--10$^4$ $M_{\odot}$.  Fluxes from the {\it Spitzer} MIPS data were used to generate the dust properties of these clusters.  We find dust masses ranging from 1--1000 $M_{\odot}$ and temperatures between 60 and 90 K.  At optical and UV wavelengths, we detected the majority of clusters at H$\alpha$, 2271 \AA, and 1528 \AA.  The free-free radii and densities range from 5--20$\times$10$^{17}$ cm and from 1500--9000 e$^-$ cm$^{-3}$.  Extinctions were derived from the radio, $JHK_S$, H$\alpha$, and UV fluxes, range from equivalent $A_{K_S}$ values of 0 to 1.29 mag (removing Galactic reddening).  
For an individual object, the maximum difference between various $A_{K_S}$ measurements is a factor of 1.6 on average (3 at worst), and considering our assumptions for calculating the extinctions and the uncertainties of the data, these factors are acceptable (see \S\S \ref{extinct} and \ref{extinct2} and Table \ref{tab3}).  More data and modeling are required to determine what causes these variations: geometry or model assumptions.  

The H$\alpha$/{\it Spitzer} morphologies show no differences in their IRAC and MIPS colors.  Half the objects have a young (i.e., spherical) morphology while the others have an older (i.e., shell) morphology.  Objects 1, 4, 6, 9, and 13 are not detected in the UV, possibly indicating larger amounts of extinction than the rest.  Objects 1 and 4 are the background objects.  Objects 9 and 13 are also those with spherical morphologies while object 6 has a shell morphology but young (3 Myr) age.  

In general, eleven of the fourteen objects in this radio-selected sample are optically-detected star clusters.  We must conclude that radio detections of point-like thermal sources do not show embedded star clusters, but the brighter free-free emitting H~{\sc ii} regions within M 33.  We propose that infrared images (e.g., MIPS 24 $\mu$m) must be used to discern which objects in a galaxy are enshrouded in their natal molecular cloud material.  These 24 $\mu$m sources must be checked with optical data (e.g., {\it UBVRI} and/or H$\alpha$) to determine those that are truly embedded.

\acknowledgements

B.B.\ and H.A.K.\ are supported by NASA grant NRA-00-01-LTSA-052.
B.B. would like to thank the {\it Spitzer} help desk for all questions concerning MIPS aperture corrections.  

Atlas Image mosaics of M 33 were obtained as part of the Two Micron All Sky Survey (2MASS), a joint project of the University of Massachusetts and the Infrared Processing and Analysis Center/California Institute of Technology, funded by the National Aeronautics and Space Administration and the Nation Science Foundation.

This research has made use of the NASA/IPAC Extragalactic Database (NED) which is operated by the Jet Propulsion Laboratory, California Institute of Technology, under contract with the National Aeronautics and Space Administration.

This work is based in part on observations made with the {\it Spitzer Space Telescope}, which is operated by the Jet Propulsion Laboratory, California Institute of Technology under a contract with NASA. Support for this work was provided by NASA through Contract Number 960785 issued by JPL/Caltech to RDG at the University of Minnesota.

GALEX is a NASA Small Explorer, launched in 2003 April. We gratefully acknowledge NASA's support for construction, operation, and science analysis of the GALEX mission, developed in cooperation with the Centre National d'Etudes Spatiales of France and the Korean Ministry of Science and Technology.

\clearpage

%\begin{figure}
%\plotone{f1a.eps}
%\plotone{f1b.eps}
%\end{figure}
\begin{figure}
\epsscale{1.0}
%\plotone{f1c.eps}
\figcaption{Near-UV, H$\alpha$, $JHK_S$, IRAC, and MIPS images of object 1.  Each image is $\sim$90\arcsec~on a side with an orientation of North up, and east left.  The radio position from \citet{johnson01} is designated by the circle in the center of the images. The radius of the circle reflects roughly the uncertainty in their position.  In the 4.5 $\mu$m image, we have overlaid the 6 cm map from \citet{duric93}.  The candidates are labeled in both the J and 3.6 $\mu$m images.  We located object 1 as the point source on the upper left of the circle.  However, the 6 cm radio source is found to that candidate's lower right.  A very faint source does exist closer to this position but is detected only at near-UV, 3.6 $\mu$m, and 4.5 $\mu$m.  We think that object 1 is a background object, possibly an AGN or QSO.  \label{m33object1}}
\end{figure}

\clearpage

%\begin{figure}
%\plotone{f2a.eps}
%\plotone{f2b.eps}
%\end{figure}

\begin{figure}
%\epsscale{0.75}
%\plotone{f2c.eps}
\figcaption{Near-UV, H$\alpha$, $JHK_S$, IRAC, and MIPS images of object 2.  Orientation and notation is the same as Figure \ref{m33object1}.  Object 2 is located in the lower left of the radio position uncertainty circle. The Wolf-Rayet star is found at the western edge of the uncertainty circle.  The 6 cm radio contours in the 4.5 $\mu$m image arise on the western side of the bipolar (see H$\alpha$) nebula.  No radio emission appears to be associated with the bright IRAC object.  
\label{m33object2}}
\end{figure}

\clearpage

%\begin{figure}
%\plotone{f3a.eps}
%\plotone{f3b.eps}
%\end{figure}

\begin{figure}
%\epsscale{0.75}
%\plotone{f3c.eps}
\figcaption{Near-UV, H$\alpha$, $JHK_S$, IRAC, and MIPS images of object 3.  Orientation and notation is the same as Figure \ref{m33object1}. The object is found in the uncertainty circle at all wavelengths and is associated with the 6 cm point source.  
\label{m33object3}}
\end{figure}

\clearpage

%\begin{figure}
%\epsscale{0.75}
%\plotone{f4a.eps}
%\plotone{f4b.eps}
%\end{figure}
\begin{figure}
%\plotone{f4c.eps}
\figcaption{Near-UV, H$\alpha$, $JHK_S$, IRAC, and MIPS images of object 4.  Orientation and notation is the same as Figure \ref{m33object1}. The object is found nearly in the center of the radio position uncertainty circle in the IRAC images only. The 6 cm emission contours are centered on this object in the 4.5 $\mu$m image.  We think that it is a background object, possibly an AGN or QSO.  
\label{m33object4}
}
\end{figure}

\clearpage

%\begin{figure}
%\epsscale{0.75}
%\plotone{f5a.eps}
%\plotone{f5b.eps}
%\end{figure}
\begin{figure}
%\plotone{f5c.eps}
\figcaption{Near-UV, H$\alpha$, $JHK_S$, IRAC, and MIPS images of object 5.  Orientation and notation is the same as Figure \ref{m33object1}.  Two objects, 5a and 5b, appear on opposite sides of the uncertainty circle.  Because the free-free emission contours connect both objects, we think that both are associated, and the combination of the two gives rise to the radio emission.  This combination of the two objects is termed object 5s and is labeled in the 8 $\mu$m image.  
\label{m33object5}}
\end{figure}

\clearpage

%\begin{figure}
%\epsscale{0.75}
%\plotone{f6a.eps}
%\plotone{f6b.eps}
%\end{figure}
\begin{figure}
%\plotone{f6c.eps}
\figcaption{Near-UV, H$\alpha$, $JHK_S$, IRAC, and MIPS images of object 6.  Orientation and notation is the same as Figure \ref{m33object1}.  A shell morphology is seen at 8 $\mu$m and 24 $\mu$m and designated 6s.  Object 6b is the brightest PAH/dust emitter, but a stellar source (object 6a) is found within the uncertainty circle.  No 6 cm emission appears to be centrally concentrated in the middle of the image.  
\label{m33object6}}
\end{figure}

\clearpage

%\begin{figure}
%\epsscale{0.75}
%\plotone{f7a.eps}
%\plotone{f7b.eps}
%\end{figure}
\begin{figure}
%\plotone{f7c.eps}
\figcaption{Near-UV, H$\alpha$, $JHK_S$, IRAC, and MIPS images of object 7.  Orientation and notation is the same as Figure \ref{m33object1}.  Object 7a is the actual star cluster, but no 6 cm emission appears to be associated with this object.  Object 7b is the only nearby 6 cm emitter and will contaminate MIPS 70 $\mu$m and 160 $\mu$m fluxes of object 7a. 
\label{m33object7}}  
\end{figure}

\clearpage

%\begin{figure}
%\epsscale{0.75}
%\plotone{f8a.eps}
%\plotone{f8b.eps}
%\end{figure}
\begin{figure}
%\plotone{f8c.eps}
\figcaption{Near-UV, H$\alpha$, $JHK_S$, IRAC, and MIPS images of object 8.  Orientation and notation is the same as Figure \ref{m33object1}.  The only cluster candidate is found within the radio position uncertainty circle, and the 6 cm radio contours are exactly located upon this object.   
\label{m33object8}}
\end{figure}

\clearpage

%\begin{figure}
%\epsscale{0.75}
%\plotone{f9a.eps}
%\plotone{f9b.eps}
%\end{figure}
\begin{figure}
%\plotone{f9c.eps}
\figcaption{Near-UV, H$\alpha$, $JHK_S$, IRAC, and MIPS images of object 9.  Orientation and notation is the same as Figure \ref{m33object1}.  The only cluster candidate is found within the radio position uncertainty circle, and the 6 cm radio contours are exactly located on top of the IRAC object. 
\label{m33object9}}
\end{figure}

\clearpage

%\begin{figure}
%\epsscale{1.0}
%\plotone{f10a.eps}
%\plotone{f10b.eps}
%\end{figure}
\begin{figure}
%\plotone{f10c.eps}
\figcaption{Near-UV, H$\alpha$, $JHK_S$, IRAC, and MIPS images of object 10.  Orientation and notation is the same as Figure \ref{m33object1}.  The large circle in the 8 $\mu$m image designates object 10s.  Since the two possible objects appear to be connected via their 6 cm emission, we think that the combination of the two (termed 10s) is the actual object.  
\label{m33object10}}
\end{figure}

\clearpage

%\begin{figure}
%\epsscale{1.0}
%\plotone{f11a.eps}
%\plotone{f11b.eps}
%\end{figure}
\begin{figure}
%\plotone{f11c.eps}
\figcaption{Near-UV, H$\alpha$, $JHK_S$, IRAC, and MIPS images of object 11.  Orientation and notation is the same as Figure \ref{m33object1}.  The large circle in the 8 $\mu$m image designates object 11s.   Object 11a is the stellar source, but the PAH/dust emission for this cluster is emanating from a shell around object 11a.  The 6 cm contours superimposed on the 4.5 $\mu$m image suggest that the free-free emission emanates from the shell of dust/PAH emission around object 11a.  
\label{m33object11}}
\end{figure}

\clearpage

%\begin{figure}
%\epsscale{1.0}
%\plotone{f12a.eps}
%\plotone{f12b.eps}
%\end{figure}
\begin{figure}
%\plotone{f12c.eps}
\figcaption{Near-UV, H$\alpha$, $JHK_S$, IRAC, and MIPS images of object 12.  Orientation and notation is the same as Figure \ref{m33object1}. Object 12 is found at the southwestern part of the uncertainty circle, is the source of the brightest 6 cm emission, and appears to have a spherical morphology. 
 \label{m33object12}}
\end{figure}

\clearpage

%\begin{figure}
%\epsscale{1.0}
%\plotone{f13a.eps}
%\plotone{f13b.eps}
%\end{figure}
\begin{figure}
%\plotone{f13c.eps}
\figcaption{Near-UV, H$\alpha$, $JHK_S$, IRAC, and MIPS images of object 13.  Orientation and notation is the same as Figure \ref{m33object1}.   This object with a spherical morphology is found at the center of the radio position uncertainty circle and is the source of the brightest 6 cm emission.  
\label{m33object13}}
\end{figure}

\clearpage

%\begin{figure}
%\epsscale{1.0}
%\plotone{f14a.eps}
%\plotone{f14b.eps}
%\end{figure}
\begin{figure}
%\plotone{f14c.eps}
\figcaption{Near-UV, H$\alpha$, $JHK_S$, IRAC, and MIPS images of object 14.  Orientation and notation is the same as Figure \ref{m33object1}.  The large circle in the 8 $\mu$m image designates object 14s.  While 14a is the stellar component, the PAH and dust emission appears to be coming from the cometary-like emission around the radio position (termed 14s).  
Some 6 cm emission appears to be associated with the stellar source, but no emission is associated with the cometary PAH/dust emission.  
\label{m33object14}}
\end{figure}

\clearpage

\begin{figure}
%\epsscale{0.8}
%\plotone{f15.eps}
\figcaption{$JHK_S$ color-color diagram of eight objects.  The legend at the upper left indicates which symbol goes with which object.  Uncertainties are 1$\sigma$ in length for all colors.  The connected crosses show the evolutionary model tracks for instantaneous star forming bursts from Starburst 99 for metallicity Z = 0.008 (solid) and for metallicity Z = 0.02 (dotted).  Crosses along the model tracks indicate ages in step of 1 Myr.  Specific ages are labeled along the tracks (0, 6, 10, and 20 Myr).  The arrow to the right shows the reddening vector for A$_{\rm V}=5$.    
\label{m33jhk}}
\end{figure}

\clearpage

\begin{figure}
%\epsscale{1.0}
%\plotone{f16.eps}
\figcaption{Spectral energy distributions of objects 2, 3, 5, and 6 from the ultraviolet to the mid-infrared.  Uncertainties for these fluxes are the size of or smaller than the symbols plotted.  Downward-pointing arrows are upper limits.  If multiple objects are designated for one position, a legend is provided to indicate which symbol goes with which object.
\label{m33sed1}}
\end{figure}

\clearpage 

\begin{figure}
%\epsscale{1.0}
%\plotone{f17.eps}
\figcaption{Spectral energy distributions of objects 7, 8, 9, and 10 from the ultraviolet to the mid-infrared.  Notation is the same as that for Figure \ref{m33sed1}.   
\label{m33sed2}}
\end{figure}

\clearpage

\begin{figure}
%\epsscale{1.0}
%\plotone{f18.eps}
\figcaption{Spectral energy distribution of objects 11, 12, 13, and 14 from the ultraviolet to the mid-infrared.  Notation is the same as that for Figure \ref{m33sed1}.
\label{m33sed3}}
\end{figure}

\clearpage

\begin{figure}
%\epsscale{1.0}
%\plotone{f19.eps}
\figcaption{IRAC color-color diagram of the M 33 radio-selected objects.  The different symbols represent the morphologies discussed in \S \ref{dismorphology}.  Note that no clear separation occurs in the IRAC colors for the different morphologies.  \label{m33iraccolor}}
\end{figure}

\clearpage

\begin{figure}
%\plotone{f20.eps}
\figcaption{MIPS color-color diagram of twelve of the M 33 objects.  The point-by-point labeling is the same as that of Figure \ref{m33iraccolor}. Note that the morphologies discussed in \S \ref{dismorphology} are not separable by their MIPS colors.  
\label{m33mipscolor}}
\end{figure}

\clearpage

\clearpage 
\begin{deluxetable}{lcccccccccc}
\rotate
\tabletypesize{\tiny}
\tablewidth{0pt}
\tablecaption{Radio-Selected Object	Positions,	$JHK_S$	Magnitudes	and	Colors, and H$\alpha$ and UV Fluxes	\label{tab1}}
\tablehead{
\colhead{Object Name} & \colhead{R.\ A.\ (J2000)} & \colhead{Decl.\ (J2000)} & \colhead{J (mag)}	& \colhead{H (mag)} & \colhead{K$_{\rm S}$ (mag)}	& \colhead{J--H (mag)} &	\colhead{H--K$_{\rm S}$ (mag)} & \colhead{F$_{\nu}({\rm H}\alpha)$ [10$^{-5}$ Jy]}	& \colhead{F$_{\nu}$(0.1528 $\mu$m) [10$^{-5}$ Jy]} & \colhead{F$_{\nu}$(0.2271 $\mu$m) [10$^{-5}$ Jy]}
}
\startdata	
%	&	1	33	2.6	&	30	46	46.1	&	14.85$\pm$0.07	&	14.35$\pm$0.06	&	14.44$\pm$0.09	&	0.504$\pm$0.09	&	-0.096$\pm$0.107	&						
1	&	1	33	02.6	&	30	46	46.1	&	\nodata	&	\nodata	&	\nodata	&	\nodata	&	\nodata	&	\nodata	&	0.85$\pm$1.73 &	0.772$\pm$2.9 \\
2	&	1	33	16.1	&	30	56	42.4	&	16.33$\pm$0.14	&	16.3$\pm$0.19	&	16.19$\pm$0.24	&	0.028$\pm$0.235	&	0.114$\pm$0.308	&	8.27$\pm$0.03	&	78.4$\pm$16.7	&	67$\pm$27	\\
3	&	1	33	16.6	&	30	52	49.9	&	15.65$\pm$0.1	&	14.81$\pm$0.08	&	14.278$\pm$0.08	&	0.847$\pm$0.124	&	0.528$\pm$0.112	&	17.8$\pm$0.01	&	23.3$\pm$9.1	&	21.8$\pm$15.4	\\
4	&	1	33	37.5	&	30	47	18.5	&	$>$17.5	&	$>$17.1	&	$>$17.6	&	\nodata	&	\nodata	&	$<$0.8	&	$<$0.08	&	$<$0.05	\\
5a	&	1	33	39.1	&	30	38	08	&	$>$17.3	&	$>$17.6	&	$>$17	&	\nodata	&	\nodata	&	1.92$\pm$0.01	&	\nodata	&	\nodata	\\
5b	&	1	33	39.3	&	30	38	02.2	&	$>$16.5	&	$>$16	&	$>$15.5	&	\nodata	&	\nodata	&	4.47$\pm$0.01	&	\nodata	&	\nodata	\\
5s	&	1	33	39.2	&	30	38	03.4	&	$>$16.5	&	$>$16	&	$>$15.5	&	\nodata	&	\nodata	&	12.7$\pm$0.02	&	62.8$\pm$14.9	&	68$\pm$27	\\
6a	&	1	33	43.6	&	30	39	05	&	16.43$\pm$0.06	&	16.12$\pm$0.18	&	15.61$\pm$0.12	&	0.31$\pm$0.19	&	0.51$\pm$0.21	&	3.68$\pm$0.02	&	$<$0.4	&	$<$0.3	\\
6b	&	1	33	44	&	30	39	13.5	&	$>$18.6	&	$>$16.9	&	$>$15.3	&	\nodata	&	\nodata	&	$<$0.5	&	$<$0.4	&	$<$0.3	\\
6s	&	1	33	44.1	&	30	39	03.5	&	$>$14.4	&	$>$16.9	&	$>$16	&	\nodata	&	\nodata	&	14.2$\pm$0.05	&	\nodata	&	\nodata	\\
7a	&	1	33	48.1	&	30	39	18.2	&	16.49$\pm$0.12	&	15.86$\pm$0.2	&	15.04$\pm$0.1	&	0.63$\pm$0.23	&	0.82$\pm$0.22	&	2.7$\pm$0.008	&	$<$0.3	&	$<$0.2	\\
7b	&	1	33	48.3	&	30	39	35.5	&	18.54$\pm$0.96	&	16.5$\pm$0.34	&	15.5$\pm$0.15	&	2.04$\pm$1.02	&	1$\pm$0.37	&	3.85$\pm$0.01	&	6.6$\pm$4.9	&	4.94$\pm$7.34	\\
8	&	1	33	59.8	&	30	32	44.2	&	$>$17.4	&	$>$16.9	&	$>$17.9	&	\nodata	&	\nodata	&	5.15$\pm$0.007	&	6.52$\pm$4.8	&	7.81$\pm$9.23	\\
9	&	1	34	00.2	&	30	40	47.5	&	16.16$\pm$0.07	&	15.4$\pm$0.08	&	14.32$\pm$0.03	&	0.76$\pm$0.1	&	1.08$\pm$0.08	&	4.67$\pm$0.01	&	$<$0.2	&	$<$0.2	\\
10a	&	1	34	02.6	&	30	38	38.7	&	$>$16	&	$>$13.8	&	$>$14	&	\nodata	&	\nodata	&	6.9$\pm$0.04	&	23.7$\pm$9.2	&	25.7$\pm$16.7	\\
10b	&	1	34	02	&	30	38	39	&	$>$16	&	$>$13.8	&	$>$14	&	\nodata	&	\nodata	&	2.73$\pm$0.02	&	$<$0.3	&	$<$0.3	\\
10s	&	1	34	02.2	&	30	38	40.8	&	$>$16	&	$>$13.8	&	$>$14	&	\nodata	&	\nodata	&	27.4$\pm$0.05	&	\nodata	&	\nodata	\\
11a	&	1	34	06.6	&	30	41	47.3	&	15.73$\pm$0.05	&	15.98$\pm$0.27	&	15.38$\pm$0.07	&	-0.25$\pm$0.28	&	0.6$\pm$0.28	&	4.98$\pm$0.02	&	102$\pm$19	&	79.9$\pm$29.5	\\
11s	&	1	34	06.6	&	30	41	47.3	&	$>$15.3	&	$>$15.6	&	$>$14.1	&	\nodata	&	\nodata	&	27.9$\pm$0.02	&	\nodata	&	\nodata	\\
12	&	1	34	13.5	&	30	34	49.3	&	$>$17.3	&	$>$15.8	&	$>$15.3	&	\nodata	&	\nodata	&	5.25$\pm$0.01	&	7.54$\pm$5.17	&	8.37$\pm$9.55	\\
13	&	1	34	17.5	&	30	33	45.8	&	$>$17.4	&	$>$16.4	&	$>$16	&	\nodata	&	\nodata	&	4.6$\pm$0.01	&	$<$0.6	&	$<$0.4	\\
14a	&	1	34	38.6	&	30	43	56.8	&	15.49$\pm$0.02	&	15.01$\pm$0.03	&	14.82$\pm$0.02	&	0.479$\pm$0.032	&	0.187$\pm$0.033	&	3.22$\pm$0.01	&	72.1$\pm$16	&	70$\pm$28	\\
14s	&	1	34	39	&	30	43	55.1	&	$>$15	&	$>$14.6	&	$>$14.5	&	\nodata	&	\nodata	&	8.21$\pm$0.01	&	\nodata	&	\nodata	\\
\enddata
\tablecomments{Units of	right ascension	are	hours, minutes, and seconds, and the units of declination are degrees, arcminutes, and arcseconds. All magnitudes and fluxes have not been dereddened. Numbers after the $\pm$ are 1$\sigma$ uncertainties.}
\end{deluxetable}
 
\clearpage 
\begin{deluxetable}{lccccccc}
\tabletypesize{\scriptsize}
\tablewidth{0pt}
\tablecaption{IRAC	and	MIPS	Fluxes	of M 33 Radio-Selected Objects \label{tab2}}
\tablehead{
\colhead{}	&	\colhead{IRAC	Ch.\	1}	&	\colhead{IRAC	Ch.\	2}	&	\colhead{IRAC	Ch.\	3}	&	\colhead{IRAC	Ch.\	4}	&	\colhead{MIPS	Ch.\	1}	&	\colhead{MIPS	Ch.\	2}	&	\colhead{MIPS	Ch.\	3}	\\
\colhead{Object Name}	&	\colhead{3.6 $\mu$m (mJy)}	&	\colhead{4.5 $\mu$m (mJy)}	&	\colhead{5.8 $\mu$m (mJy)}	&	\colhead{8 $\mu$m (mJy)}	&	\colhead{24 $\mu$m (mJy)}	&	\colhead{70 $\mu$m (mJy)}	&	\colhead{160 $\mu$m (mJy)}}

\startdata
%1	&	0.611$\pm$0.036	&	0.366$\pm$0.021	&	0.222$\pm$0.014	&	0.195$\pm$0.012	&	$<$0.09	&	$<$2	&	$<$70	\\
2	&	0.383$\pm$0.032	&	0.374$\pm$0.031	&	1.16$\pm$0.1	&	3.2$\pm$0.3	&	57$\pm$1	&	1190$\pm$270	&	940$\pm$200	\\
3	&	6.22$\pm$0.51	&	11.6$\pm$1	&	32.8$\pm$2.7	&	92.7$\pm$7.6	&	1700$\pm$200	&	7340$\pm$1810	&	4150$\pm$880	\\
4	&	0.182$\pm$0.016	&	0.228$\pm$0.02	&	0.221$\pm$0.02	&	0.259$\pm$0.023	&	$<$0.08	&	$<$5	&	$<$200	\\
5a	&	0.662$\pm$0.055	&	0.557$\pm$0.047	&	4.03$\pm$0.34	&	12.1$\pm$1	&	\nodata	&	\nodata	&	\nodata	\\
5b	&	0.996$\pm$0.083	&	0.653$\pm$0.055	&	4.13$\pm$0.34	&	11.6$\pm$1	&	\nodata	&	\nodata	&	\nodata	\\

5s	&	2.97$\pm$0.24	&	2.01$\pm$0.17	&	13.7$\pm$1.4	&	39.7$\pm$3.2	&	142$\pm$21	&	1310$\pm$320	&	1360$\pm$340	\\
6a	&	0.155$\pm$0.016	&	0.152$\pm$0.014	&	0.293$\pm$0.037	&	0.981$\pm$0.11	&	\nodata	&	\nodata	&	\nodata	\\
6b	&	0.529$\pm$0.045	&	0.417$\pm$0.035	&	2.11$\pm$0.18	&	6.17$\pm$0.51	&	\nodata	&	\nodata	&	\nodata	\\
6s	&	3.54$\pm$0.3	&	2.66$\pm$0.23	&	17.7$\pm$1.4	&	51.4$\pm$4.2	&	101$\pm$10	&	1810$\pm$450	&	2080$\pm$480	\\
7a	&	0.79$\pm$0.066	&	0.508$\pm$0.04	&	2.59$\pm$0.21	&	8.11$\pm$0.68	&	\nodata	&	\nodata	&	\nodata	\\
7b	&	1.53$\pm$0.13	&	1.24$\pm$0.1	&	8.17$\pm$0.68	&	24.1$\pm$2	&	194$\pm$43	&	1120$\pm$280	&	1210$\pm$320	\\
8	&	0.749$\pm$0.062	&	0.585$\pm$0.049	&	3.44$\pm$0.28	&	9.75$\pm$0.8	&	42.1$\pm$7.9	&	1730$\pm$370	&	2550$\pm$129	\\
9	&	5.85$\pm$0.48	&	6.12$\pm$0.51	&	31.7$\pm$2.7	&	99.4$\pm$0.82	&	829$\pm$182	&	426$\pm$95	&	3230$\pm$640	\\
10a	&	0.825$\pm$0.069	&	0.691$\pm$0.058	&	4.59$\pm$0.38	&	12.6$\pm$1	&	\nodata	&	\nodata	&	\nodata	\\
10b	&	1.58$\pm$0.13	&	1.29$\pm$0.11	&	8.24$\pm$0.68	&	24.6$\pm$2	&	\nodata	&	\nodata	&	\nodata	\\
10s	&	6.08$\pm$0.51	&	4.97$\pm$0.41	&	34$\pm$3	&	93.2$\pm$7.6	&	254$\pm$29	&	5560$\pm$1050	&	3960$\pm$850	\\
11a	&	0.32$\pm$0.027	&	0.339$\pm$0.028	&	$<$0.04	&	$<$0.1	&	\nodata	&	\nodata	&	\nodata	\\
11s	&	2.74$\pm$0.23	&	2.12$\pm$0.17	&	8.6$\pm$0.7	&	24.4$\pm$2	&	75.1$\pm$7.9	&	4120$\pm$1000	&	1790$\pm$410	\\
12	&	1.04$\pm$0.08	&	0.852$\pm$0.07	&	4.07$\pm$0.34	&	11$\pm$0.8	&	108$\pm$11	&	1260$\pm$260	&	3470$\pm$750	\\
13	&	1$\pm$0.08	&	0.745$\pm$0.062	&	4.72$\pm$0.39	&	12.9$\pm$1.1	&	93.9$\pm$17	&	1580$\pm$390	&	3600$\pm$780	\\
14a	&	0.529$\pm$0.044	&	0.318$\pm$0.027	&	0.908$\pm$0.076	&	2.3$\pm$0.2	&	\nodata	&	\nodata	&	\nodata	\\
14s	&	1.02$\pm$0.08	&	0.654$\pm$0.055	&	2.94$\pm$0.24	&	7.42$\pm$0.61	&	41.1$\pm$5.5	&	1620$\pm$260	&	1280$\pm$310	\\
\enddata
\end{deluxetable}

\clearpage
\begin{deluxetable}{lcccccccc}
\rotate 
\tabletypesize{\scriptsize}
\tablewidth{0pt}																								
\tablecaption{Extinctions derived from JHK$_{\rm S}$, H$\alpha$, UV, and Radio Data along with Derived Stellar Ages, Stellar Masses, Dust Temperatures, and Dust Masses \label{tab3}}
\tablehead{
\colhead{Object	Name} & \colhead{A$_{{\rm K}_{\rm S}}$ (mag)} & \colhead{A$_{{\rm H}\alpha}$ (mag)} & \colhead{A$_{0.2271}$ (mag)} & \colhead{A$_{0.1528}$ (mag)} & \colhead{Stellar Age (Myr)}	& \colhead{Log(Stellar Mass) [Log(M$_{\odot}$)]} & \colhead{Dust Temperature (K)} & \colhead{Dust Mass (M$_{\odot}$)} 
}
\startdata
1%	&	\nodata	&	\nodata	&	\nodata	&	\nodata	&	\nodata	&	\nodata	&	\nodata	&	\nodata	\\
2	&	\nodata	&	0.72	&	\nodata	&	\nodata	&	\nodata	&	\nodata	&	64(35)	&	12(330)	\\
3	&	0.84	&	2.01	&	3.88	&	4.2	&	7$^{+1}_{-2}$	&	3.5	&	87(44)	&	31(810)	\\
4%	&	\nodata	&	\nodata	&	\nodata	&	\nodata	&	\nodata	&	\nodata	&	\nodata	&	\nodata	\\
5a	&	\nodata	&	2.57	&	\nodata	&	\nodata	&	\nodata	&	\nodata	&	\nodata	&	\nodata	\\
5b	&	\nodata	&	1.65	&	\nodata	&	\nodata	&	\nodata	&	\nodata	&	\nodata	&	\nodata	\\
5s	&	\nodata	&	0.51	&	\nodata	&	\nodata	&	\nodata	&	\nodata	&	66	&	12	\\
6a	&	0.23	&	1.8	&	\nodata	&	\nodata	&	3$^{+1}_{-3}$	&	3	&	\nodata	&	\nodata	\\
6b	&	\nodata	&	\nodata	&	\nodata	&	\nodata	&	\nodata	&	\nodata	&	\nodata	&	\nodata	\\
6s	&	\nodata	&	0.33	&	\nodata	&	\nodata	&	\nodata	&	\nodata	&	61	&	22	\\
7a	&	0.54	&	1.06	&	\nodata	&	\nodata	&	2$\pm$2	&	3.15	&	\nodata	&	\nodata	\\
7b	&	1.31	&	0.68	&	4.18	&	4.6	&	10$_{-10}^{+1000}$	&	3.14	&	73	&	7.6	\\
8	&	\nodata	&	0.36	&	\nodata	&	\nodata	&	\nodata	&	\nodata	&	66	&	16	\\
9	&	0.68	&	3.4	&	\nodata	&	\nodata	&	$<$6	&	3.4	&	92	&	1.5	\\
10a	&	\nodata	&	2.45	&	\nodata	&	\nodata	&	\nodata	&	\nodata	&	\nodata	&	\nodata	\\
10b	&	\nodata	&	3.46	&	\nodata	&	\nodata	&	\nodata	&	\nodata	&	\nodata	&	\nodata	\\
10s	&	\nodata	&	0.96	&	\nodata	&	\nodata	&	\nodata	&	\nodata	&	65(40)	&	55(980)	\\
11a	&	0.79	&	1.65	&	0.686	&	1.9	&	7$^{+2}_{-3}$ &	3.2	&	\nodata	&	\nodata	\\
11s	&	\nodata	&	0	&	\nodata	&	\nodata	&	\nodata	&	\nodata	&	57	&	64	\\
12	&	\nodata	&	1.6	&	\nodata	&	\nodata	&	\nodata	&	\nodata	&	61	&	15	\\
13	&	\nodata	&	1.44	&	\nodata	&	\nodata	&	\nodata	&	\nodata	&	58	&	23	\\
14a	&	0.3	&	1.76	&	0.15	&	0.43	&	7.5$_{-0.5}^{+1000}$	&	3.4	&	\nodata	&	\nodata	\\
14s	&	\nodata	&	0.75	&	\nodata	&	\nodata	&	\nodata	&	\nodata	&	56	&	27	\\
\enddata 
\tablecomments{Numbers	after the $\pm$	are	1$\sigma$ uncertainties. Extinctions recorded in this Table do not have the Galactic reddening	component removed (an $A_{K_S}$ of 0.015 mag).  $A_{K_S}$ is derived from the distance between the Starburst	99  model $JHK_S$ colors and the reddened $JHK_S$ colors. $A_{H\alpha}$ is derived	from the radio and H$\alpha$ fluxes used in the	equation from \citet{caplan86}. $A_{0.1528 \micron}$ and $A_{0.2271 \micron}$ were derived	by comparing Starburst 99 fluxes at these wavelengths with those measured (see Table \ref{tab1} for UV fluxes). Dust temperature and dust masses found in the parentheses are the second results from the two blackbody fits.}
\end{deluxetable}

\end{document}